# Modulation of Spin Accumulation by Nanoscale Confinement using Electromigration in a Metallic Lateral Spin Valve

**Emmanouil Masourakis[1], Libe Arzubiaga[1], Goran Mihajlović[2], Estitxu Villamor[1], Roger Llopis[1], Fèlix Casanova[1, 3,*], Luis E. Hueso[1,3]**

[1]CIC nanoGUNE, 20018 Donostia—San Sebastian, Basque Country, Spain

[2] San Jose Research Center, HGST, a Western Digital Company, San Jose, California 95135, USA

[3]IKERBASQUE, Basque Foundation for Science, 48011 Bilbao, Basque Country, Spain

*E-mail: f.casanova@nanogune.eu

**Abstract**

We study spin transport in lateral spin valves with constricted channels. Using electromigration, we modulate the spin accumulation by continuously varying the width of the non-magnetic channel at a single location. By fitting the non-local spin signal data as a function of the non-magnetic channel resistance, we extract all the relevant parameters regarding spin transport from a single device. Simulations show that constricting the channel blocks the diffusion of the accumulated spins rather than causing spin flipping. This result could be used to improve the design of future spintronic devices devoted to information processing.

## 1. Introduction

Spintronics is an alternative to mainstream electronics that makes use of the spin of the electron to store and process information [1]. Prominent examples of applied spintronic devices are hard-disk read heads and magnetic random access memories (MRAM), which make use of the giant magnetoresistance [2, 3] and tunnel magnetoresistance [4, 5, 6] effects. Further progress could be achieved with the use in devices of pure spin currents (*i.e.*, a flow of spin angular momentum without being accompanied by a charge current), which are an essential ingredient in an envisioned spin-only circuit [7].

Lateral spin valves (LSVs) are basic spintronic devices that create, transport and detect pure spin currents, being an attractive means to study both spin transport as well as spin injection properties in different materials [8, 9, 10, 11, 12, 13, 14, 15, 16, 17] [18]. LSVs consist of two ferromagnetic (FM) electrodes, used to inject and detect pure spin currents, bridged by a non-magnetic (NM) channel, which transports the injected spin current (see Figure. 1(a)). Confinement effects related to such nanostructures play an important role in the spin transport, as they might be used to enhance the magnitude of the spin currents [19, 12] or, on the contrary, could introduce additional sources of spin relaxation [15, 11]. Modulation of spin currents is an elusive objective of spintronics [20, 21, 22] and investigating its relationship to confinement effects could help with the development of future applications.

In this manuscript, we study the role of nanoscale confinement effects by continuously tuning the size of a constriction in the NM channel of a LSV. Using electromigration (EM), we modulate the spin accumulation by actively varying the width of the NM channel at a single location. The control of the confinement effect allows us to extract the spin diffusion length of the NM ($\lambda_N$), and the spin polarization of the FM ($\alpha_F$) using a single LSV device. Using numerical calculations of the Valet-Fert model [23] in 3D, we then simulate the spin accumulation and find that the effect of the constriction is to confine the spin population without increasing spin-flip scattering. The injected spin population is contained close to the injection interfaces resulting in decreased spin accumulation for the non-local configuration.

## 2. Methods

*2.1 Fabrication and Non-Local Measurement*

LSV devices were fabricated on silicon (Si) substrates covered with 150 nm of silicon dioxide ($SiO_2$) by a two-step electron-beam lithography, ultra-high vacuum evaporation and lift-off process. Further details regarding the fabrication procedure can be found in some of our previous publications [11, 24]. In this particular case (see Figure 1(a)), the FM electrodes are made of 30-nm-thick $Ni_{80}Fe_{20}$ (permalloy, Py) and were given different widths (125 and 80 nm) in order to have different switching fields. The NM channel is a 40-nm-thick and 130-nm-wide Cu strip, with a constriction that reaches 70 nm in width at its narrowest point. The distance *L* between the Py electrodes is 500 nm. All electrical measurements were performed in a liquid-He cryostat at 10 K, applying a "DC-reversal" technique with typical current values of *I*=100 µA [13]. EM process was performed in a 2-point configuration [25, 26], while the FM electrodes where left at a floating voltage. The resistivity of Py was measured in a different nanowire with the same nominal geometry and found to be $\rho_{Py}$ = 28.2 µΩcm at 10 K.

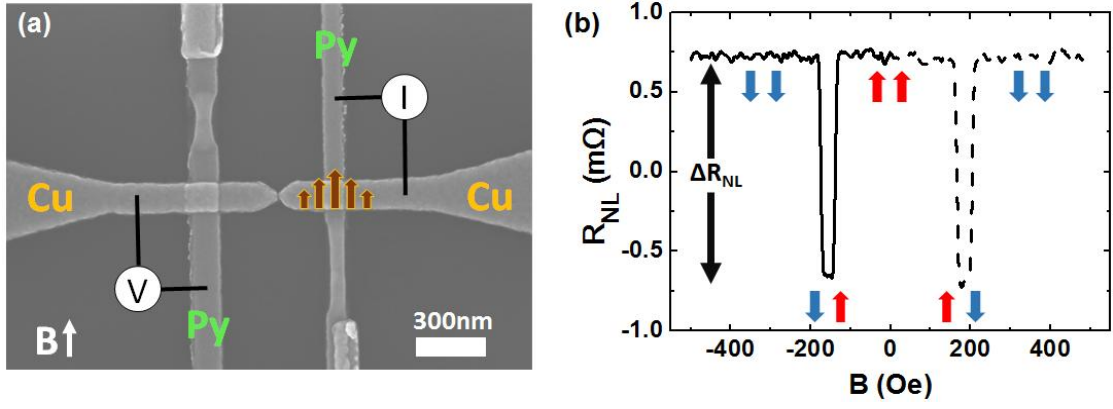

Figure 1. (a) Scanning electron micrograph of a LSV. In the non-local measurement a spin-polarized current is injected from one FM electrode to the NM channel; the accumulated spin population causes a voltage detected by a second FM electrode. The direction of the applied magnetic field (*B*) is shown. (b) Non-local resistance, measured at 10 K, as a function of *B*. Solid (dashed) line indicates the decreasing (increasing) direction of *B*. The arrows represent the relative magnetization of the Py electrodes, indicating the parallel and anti-parallel states. Spin signal is tagged as $\Delta R_{NL}$.

Typically, spin valve measurements involve the detection of a change in resistance $\Delta R_{NL}$ ($R_{NL}=V_{detected}/I_{injected}$), between the states of parallel and antiparallel orientation of the FM electrodes. This quantity is a direct representation of the spin accumulation in the NM channel at the FM/NM detector interface. The spin signal $\Delta R_{NL}$ is typically obtained from measurements of $R_{NL}$ as a function of the magnetic field *B*, as shown in Figure 1(b). By solving the one-dimensional (1D) spin diffusion equation for transparent contacts, the spin signal can be expressed as follows [9, 27]:

$$\Delta R_{NL} = \frac{4R_N \left[\alpha_F \frac{R_{SF}}{R_{SN}}\right]^2 e^{-\frac{L}{\lambda_N}}}{\left[1+\frac{2R_{SF}}{R_{SN}}\right]^2 - e^{-\frac{2L}{\lambda_N}}} \qquad (1)$$

where *L* is the edge-to-edge distance between the FM electrodes, $\alpha_F$ is the spin polarization of the FM, $R_{SN} = \lambda_N \rho_N / t_N w_N$ and $R_{SF} = \lambda_F \rho_F / w_N w_F (1-\alpha_F^2)$ are the spin resistances, $\lambda_{N,F}$ the spin diffusion lengths, $w_{N,F}$ the widths and $\rho_{N,F}$ the resistivities of NM and FM, respectively. $t_N$ is the thickness of the NM.

*2.2 Electromigration*

EM is basically the transformation and reallocation of metallic grains due to a large current density. Here, it offers the opportunity to examine the behavior of diffusing spins using the same injection and detection conditions while changing only the path of spin diffusion. During EM, a steadily increasing voltage is applied across the Cu nanowire, while the resistance is continuously measured (see Figure 2). An abrupt increase in resistance (end of voltage ramps in Figure 2) indicates a transformation of the Cu channel and signals an active feedback mechanism to restart the voltage ramp. The feedback allows us to constrict the Cu channel in small, gradual steps and observe the effect of very slight changes of channel resistance on the spin signal (see Figure 3).

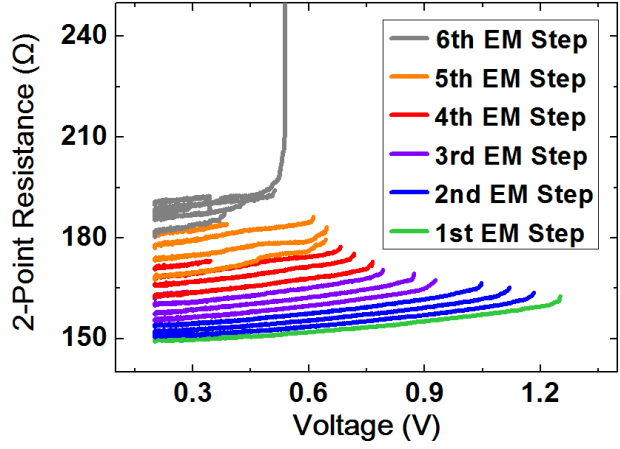

Figure 2. Electromigration of the Cu channel of a LSV. The voltage is ramped up and the 2-point resistance is monitored. When an abrupt increase in the 2-point resistance is detected, a feedback mechanism resets the voltage to the starting value. Controllable and gradual increases of the 2-point resistance allowed us to stop after several ramps and re-measure $\Delta R_{NL}$.

The EM causes a reduction in the cross sectional area which can influence both the resistivity and the spin resistance of the channel. After each EM step, the $R_{NL}$ versus $B$ measurement was repeated (see Figure 3(a)). Correspondingly, a 4-point resistance measurement of the Cu channel was performed. As expected, for each increase in the resistance of the Cu channel, $\Delta R_{NL}$ decreased (Figure 3(b)).

## 3. Results

### 3.1 Extraction of Spin Diffusion Parameters

The evolution of the spin signal is modeled using the 1D spin diffusion model. Adapting Equation 1 to our geometry we obtain:

$$\Delta R_{NL} = \frac{4\,\alpha_F^2\,\lambda_{Cu}\,\frac{R_{Cu}}{L}}{\left(2 + \lambda_{Cu}\,R_{Cu}\,\frac{w_{Py}\,w_{Cu}(1-\alpha_F^2)}{\lambda_{Py}\,L\,\rho_{Py}}\right)^2 e^{\frac{L}{\lambda_{Cu}}} - \left(\lambda_{Cu}\,R_{Cu}\,\frac{w_{Py}w_{Cu}(1-\alpha_F^2)}{\lambda_{Py}\,L\,\rho_{Py}}\right)^2 e^{-\frac{L}{\lambda_{Cu}}}}$$

(2)

where $R_{Cu}$ represents the 4-point resistance of the Cu channel. By taking the dimensions obtained from scanning electron microscopy (SEM), the measured $\rho_{Py}$ and assuming $\lambda_{Py} = 5$ nm [28, 29], we can accurately fit the measured $\Delta R_{NL}$ to Equation 2 (see red line in Figure 3(b)) and obtain $\alpha_{Py} = 42\% \pm 2.3\%$ and $\lambda_{Cu}=454$ nm $\pm$ 167 nm. The results of the fitting are in good agreement with previous work using similar Cu thickness [8, 11].

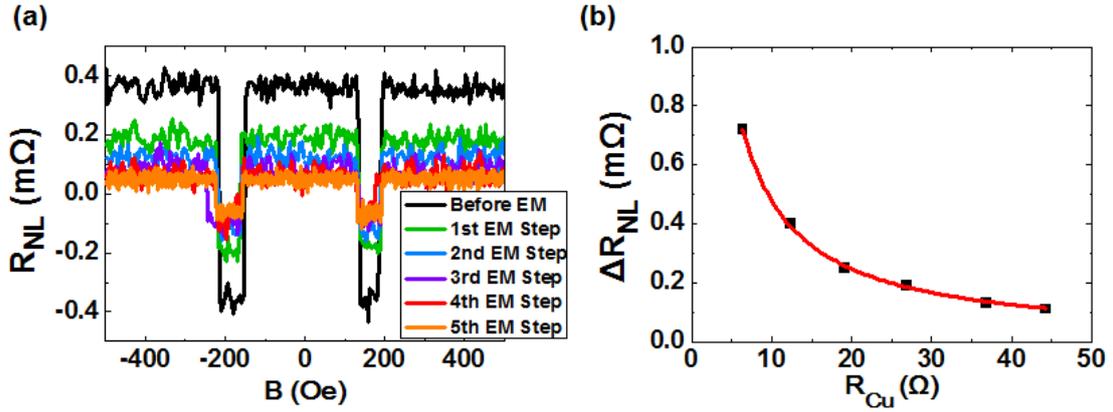

Figure 3. (a) Non-local resistance, measured at 10 K, as a function of the magnetic field after each electromigration step of the Cu channel. (b) Spin signal, obtained from the measurements in (a), as a function of the resulting 4-point Cu channel resistance after each electromigration step. Red line is a fit of the data to the 1D model given by Equation 2.

To extract such information, usually a series of devices with varying $L$ are needed [8, 10, 11, 12, 14, 17, 18, 24], while in the current approach, the relevant parameters can be extracted from a single device. Hanle measurements, which can be used to extract the same information using a single device [15], are very sensitive to device details and its liability has been recently put into question [30, 31, 32, 33]. Non-local measurements using EM on a single device provide a clear advantage, as sample-to-sample variations do not play a role.

Taking into account that a resistivity increase is expected at the constriction due to the reduced wire dimensions [34] and that spin relaxation in Cu is dominated by the Elliott-Yafet (EY) mechanism [11, 14], an increase in momentum scattering events should enhance the spin-flip scattering and decrease the spin diffusion length appropriately. For this reason, extracting a single spin diffusion length for our channel with a similar value to an unconstricted Cu channel [8, 11] is rather surprising. To further understand the effect of the constriction, we proceeded to simulate the spin accumulation and associated spin signal with numerical calculations of the Valet-Fert model [23] in 3D using the Spinflow3D software [36].

*3.2 Modelling and Simulations*

To calculate the spin accumulation within our device, we require modeling the device accurately. The channel is segmented as shown in Figure 4 and the 4-point channel resistance is assumed to be the series resistance of the segments. Resistivity measurements on independent samples reveal that the edge section resistivity is 2.2 µΩcm [11]. The resistivity of the constricted section is assumed to vary linearly with the channel width according to $w(x) = w_e + \frac{w_c - w_e}{L_c} x$ where $w$ is width, $L$ is length and the subscripts $e$ and $c$ represent the edge and constriction sections, respectively.

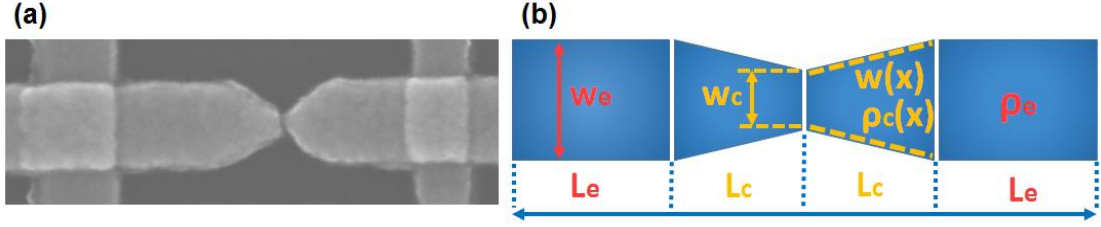

Figure 4. (a) SEM image of 500-nm-long constricted channel. Image was taken after measurements and constriction was already electromigrated. (b) Model of segmentation of channel for analysis.

The total (4-point) channel resistance $R_{tot}$ can therefore be expressed as:

$$R_{tot} = 2R_e + 2R_c = 2\frac{\rho_e L_e}{tw_e} + 2\int_0^{L_c} \frac{\rho_c(x)}{tw(x)}dx \qquad (3)$$

where $R$ is the resistance, $\rho$ is the resistivity, $t$ is the thickness and $dx$ represents infinitesimally small steps in the length axis, $x$. The premise here is that the resistivity varies linearly as a function of the channel width. Size-dependent resistivity has been clearly demonstrated in Cu nanowires [37, 38, 34] and, as we will see below, agrees perfectly with our results. The width-dependent resistivity throughout the channel can be expressed as $\rho_c(x) = A + B/w(x)$, where $A$ and $B$ are constants. Using the boundary conditions $\rho_c(0) = \rho_e = 2.2$ μΩcm and $R_{tot} = 6.4$ Ω (from our measurements), we are able to solve for $A$ and $B$ and implement a complete parametrization of the device. Equation 3 allows the channel resistance measured after each EM step to be related to a specific size of constriction (see Table I).

Table I. Measured 4-point channel resistance and corresponding constriction sizes calculated for each EM step.

| EM Stage | Measured $R_{Cu}$ (Ω) | Calculated $w_c$ (nm) |
|---|---|---|
| Before EM | 6.4 | 70 |
| EM Step 1 | 12.4 | 43.4 |
| EM Step 2 | 19.1 | 31.1 |
| EM Step 3 | 26.9 | 23.7 |
| EM Step 4 | 36.9 | 18.3 |
| EM Step 5 | 44.3 | 15.8 |

Before EM, the smallest dimension of the Cu channel is its thickness (40 nm), which can be presumed to define the average grain size [38]. As the width approaches the smallest dimension of the wire, a local reconfiguration of the grain size and also the grain boundaries will induce an increase in resistivity. Following the above model, this variation is plotted in Figure 5(a).

Implementation of our model allows the simulation of the spin signal for a variable resistivity. Using the extracted $\lambda_{Cu}$ and the calculated values of $w_c$, the spin signal is simulated for each EM step. As can be seen in Figure 5(b), the simulated values are in very good agreement with the experimental results, indicating the model propriety. This excellent agreement confirms that, despite the local increase of $\rho$ by a factor of 40, the effective $\lambda_{Cu}$ of the channel is similar to that of un-constricted Cu [11], showcasing a very interesting phenomenon.

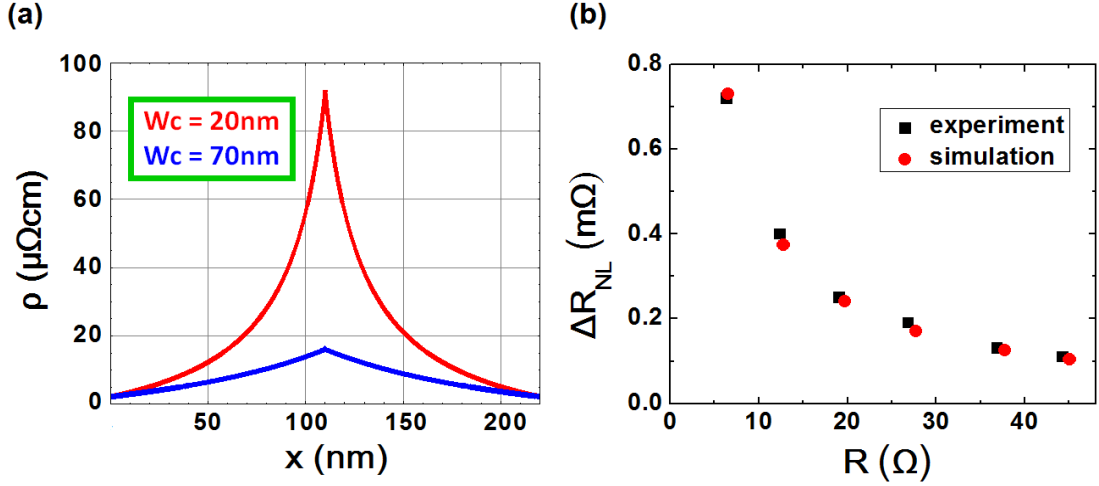

Figure 5: (a) Variation of resistivity in a 20-nm-wide (red) and 70-nm-wide (blue) constricted region. (b) Experimental data (black squares) and simulated spin signal using a 3D model with $\lambda_N$=451nm and the variable $\rho$ model described above (red circles). An excellent agreement is obtained.

## 4. Discussion

To understand this observation, we consider the two main effects of the constriction in the channel: the increase of the resistivity and the decrease of the cross section. The first effect is related to scattering within the wire, while the second one increases the spin resistance, creating a bottleneck which blocks the injected spins from getting to the detector side. The large effective $\lambda_{Cu}$ observed implies that a significant increase in scattering does not take place and the main effect of the constriction is to decrease the spin signal by simply blocking the spin diffusion. Hence, even though only a small number of spins cross the constriction, the ones which cross it do not experience a significantly larger number of scattering events and the diffusion is effectively determined by the unconstricted part of the channel.

To visualize the exact effect of the constriction we show the spin accumulation voltage $V_S$ [39] along a constricted channel for various constriction sizes. In Figure 6 we plot $V_S$ in the non-local configuration while the device is in the parallel state.

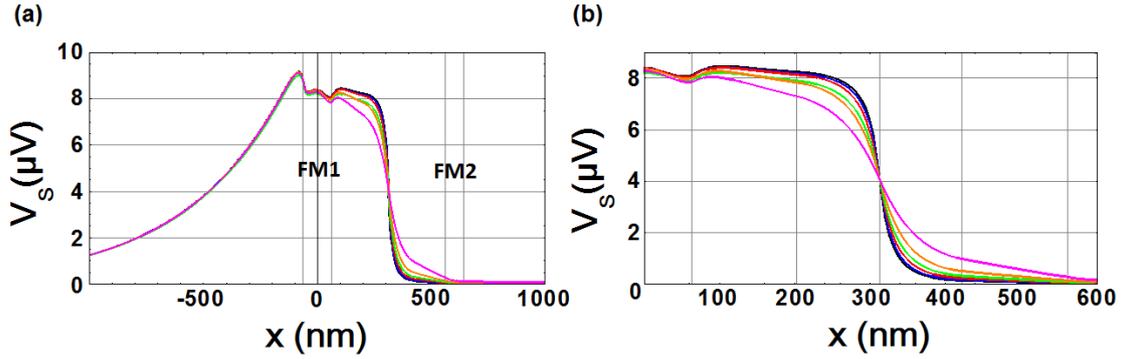

Figure 6: (a) Spin accumulation voltage due to current injection at FM1 (x=0) for different constriction sizes: 70 nm (pink), 43.3 nm (orange), 31.1 nm (green), 23.7 nm (red), 18.3 nm (blue) and 15.6 nm (black). (b) Zoom of (a) at the point of constriction.

This simulation allows the visualization of the resulting spin accumulation at each point throughout the NM channel. On the left side of the injection point we observe that the accumulated spins decay exponentially as expected for an un-constricted channel. On the right side, we see the effect of the constriction which results in a different distribution of the spin population.

On the constricted side, the spin accumulation is maintained fairly constant until the point of the constriction, resulting in an increased spin accumulation compared to its un-constricted counterpart. In addition, the spin accumulation increases when decreasing the constriction size. The decrease of the signal at the constriction signifies a more equilibrated spin population in that region, although, as discussed above, this is due to a bottleneck of the diffusing spins rather than to an increase in the spin-flip scattering. Interestingly, on the right side of the constriction (see zoom in Figure 6(b)), $V_s$ still decays exponentially, evidencing that the spin diffusion is mostly determined by $\lambda_{Cu}$ of the edge section. The simulation thus confirms that the role of the constriction is to confine and not decohere the injected spins. The constriction uses the confinement of the spin populations to modulate the spin signal. Despite the reduced spin accumulation present at the detector side of the constriction, the injector side showed a considerable increase of spin accumulation.

## 5. Conclusion

To summarize, we can modulate the spin resistance of a spin transport channel in a LSV using electromigration and dynamically probe the spin accumulation in a non-local configuration. Spin diffusion occurs in accordance to the 1D spin diffusion model, which can be used to extract the spin transport properties using a single device. Careful modeling has been used to extract the exact size of the constriction during EM. Using a numerical 3D spin diffusion model, we can visualize the effect of the constrictions on spin accumulation. Shrinking of the channel width at a given point causes a bottleneck effect, decoupling the two Py/Cu interfaces. In this light, we believe this method to provide unique and exciting prospects for modulation and analysis of spin currents in numerous spintronic devices.

**Acknowledgments**

This work was supported by the European Union 7th Framework Programme under the Marie Curie Actions (264034-QNET) and the European Research Council (257654-SPINTROS), and by the Spanish MINECO under Project No. MAT2012-37638. L.A. and E.V. thank the Basque Government for a PhD fellowship (Grants No. BFI-2009-160 and BFI-2010-163).